# Effectiveness of regional diffusion MRI measures in distinguishing multiple sclerosis abnormalities within the cervical spinal cord

Haykel Snoussi 1, 2
Julien Cohen-Adad 3, 4, 5
Benoît Combès 1
Élise Bannier 1, 6
Slimane Tounekti 7
Anne Kerbrat 8
Christian Barillot 1, *
Emmanuel Caruyer 1, *

**1:** Univ Rennes, CNRS, Inria, Inserm, IRISA UMR 6074, Empenn ERL U 1228, Rennes, France
**2:** Department of Radiology, Boston Children's Hospital, Harvard Medical School, Boston, Massachusetts, USA
**3:** NeuroPoly Lab, Institute of Biomedical Engineering, Polytechnique Montreal, Montreal, Quebec Canada
**4**: Functional Neuroimaging Unit, CRIUGM, Université de Montréal, Montréal, Quebec Canada
**5:** Mila - Quebec AI Institute, Montréal, Quebec Canada
**6:** Department of Radiology, Rennes University Hospital, Rennes, France
**7:** Department of Radiology, Thomas Jefferson University, Philadelphia, PA, USA
**8:** Departement of Neurology, Rennes University Hospital, Rennes, France
* Equal Contribution

**Correspondence:**
Haykel Snoussi, (dr.haykel.snoussi@gmail.com) Department of Radiology, Boston Children's Hospital, Harvard Medical School, Boston, Massachusetts, USA

Emmanuel Caruyer (emmanuel.caruyer@irisa.fr), Univ Rennes, CNRS, Inria, Inserm, IRISA UMR 6074, Empenn ERL U 1228, Rennes, F-35000

**Haykel Snoussi** was partially funded by the EMISEP PHRC, the Brittany region, and a MITACS-Inria Globalink travel grant. The acquisition of MRI data was supported by the Neurinfo MRI research facility at the University of Rennes 1. Neurinfo is granted by the European Union (FEDER), the French State, the Brittany Council, Rennes Metropole, Inria, Inserm, and the University Hospital of Rennes 1.

# Effectiveness of regional diffusion MRI measures in distinguishing multiple sclerosis abnormalities within the cervical spinal cord

*Abstract*

*Introduction*: Multiple sclerosis is an inflammatory disorder of the central nervous system. While conventional MRI is widely used for multiple sclerosis diagnosis and clinical follow-up, quantitative MRI has the potential to provide valuable intrinsic values of tissue properties that can enhance accuracy. In this study, we investigate the efficacy of diffusion MRI in distinguishing multiple sclerosis lesions within the cervical spinal cord, using a combination of metrics extracted from diffusion tensor imaging and Ball-and-Stick models.
*Methods*: We analyzed spinal cord data acquired from multiple hospitals and extracted average diffusion MRI metrics per vertebral level using a collection of image processing methods and an atlas-based approach. We then performed a statistical analysis to evaluate the feasibility of these metrics for detecting lesions, exploring the usefulness of combining different metrics to improve accuracy.
*Results*: Our study demonstrates the sensitivity of each metric to underlying microstructure changes in multiple sclerosis patients. We show that selecting a specific subset of metrics, which provide complementary information, significantly improves the prediction score of lesion presence in the cervical spinal cord. Furthermore, the Ball-and-Stick model has the potential to provide novel information about the microstructure of damaged tissue.
*Conclusion*: Our results suggest that diffusion measures, particularly combined measures, are sensitive in discriminating abnormal from healthy cervical vertebral levels in patients. This information could aid in improving multiple sclerosis diagnosis and clinical follow-up. Our study highlights the potential of the Ball-and-Stick model in providing additional insights into the microstructure of the damaged tissue.

## 1. Introduction

Multiple sclerosis (MS) is a chronic and debilitating disease of the central nervous system (CNS) that affects millions of people worldwide. MS is characterized by inflammation, demyelination, and neurodegeneration of the CNS, leading to a wide range of symptoms such as visual disturbances, sensory abnormalities, motor dysfunction, and cognitive impairment (Inglese & Bester, 2010; Wheeler-Kingshott et al., 2014). The spinal cord, particularly the cervical region, is frequently affected by MS and plays a crucial role in many of the symptoms experienced by patients (Bot & Barkhof, 2009; Stroman et al., 2014). In particular, spinal cord lesions are common in the cervical region and strongly associated with disability and functional deficits in MS patients (By et al., 2017). However, conventional magnetic resonance imaging (MRI) techniques have limited sensitivity and specificity for detecting and quantifying spinal cord lesions and their underlying pathology. More accurate and sensitive detection of spinal cord lesions can aid in early diagnosis, monitoring disease progression, and evaluating treatment efficacy. Additionally, improving our ability to detect spinal cord lesions in MS is important not only for the clinical management of patients but also for advancing our understanding of the disease. Diffusion-weighted MRI has emerged as a promising tool for assessing tissue microstructure and pathology in MS. This modality is sensitive to the microscopic movement of water molecules in biological tissues, which can be used to derive various metrics related to tissue diffusivity and microstructure. Diffusion tensor imaging (DTI) is a commonly used diffusion MRI technique that can quantify the diffusion of water molecules along different directions and estimate the orientation and integrity of white matter tracts in the CNS.

The use of diffusion MRI for detecting and identifying spinal cord lesions in MS still needs further investigation to extract new insights and underlying information, despite several technical and methodological challenges. For example, a few limited studies have investigated advanced diffusion models such as Diffusion Kurtosis Imaging and Neurite Orientation Dispersion and Density Imaging (By et al., 2017; Grussu et al., 2015). The application of these multi-compartment models in the spinal cord requires a high b-value and a high number of diffusion encoding directions, resulting in clinically unfeasible scan times. In addition, although their important findings, the data used is very limited in terms of the number of subjects and the volume of regions in which the diffusion metrics were quantified.

In this study, we aimed to investigate the effectiveness of regional diffusion MRI measures in distinguishing MS abnormalities and address some of these challenges by using an adapted diffusion MRI model, the Ball-and-Stick model, to assess the sensitivity and specificity of diffusion metrics in the cervical spinal cord of MS patients. The Ball-and-Stick model was chosen because it offers the advantage of identifying and separating crossing fibers, providing more detailed microstructural information compared to DTI. Our main objective was to explore the usefulness of combining different metrics derived from these two models to improve their sensitivity associated with the presence of MS lesions. We employed an atlas-based and a collection of image-processing approaches to quantify diffusion metrics at different vertebral levels within the cervical spinal cord and performed a statistical analysis to detect the presence of MS lesions.

## 2. Materials and Methods

### 2.1. Multiple sclerosis patients and healthy volunteers

This multicenter study includes 82 participants: 29 healthy volunteers (mean age = 32.83±7.13, 18F/11M) and 53 MS patients (mean age = 32.58±6.25, 34F/19M). All subjects were recruited in accordance with the approval of the local research ethics committee (EMISEP PHRC project [1]) and provided informed written consent. The MS patients included in this study were early relapsing remitting MS patients, with a median Expanded Disability Status Scale (EDSS) score of 1.0 (range [0, 2.5]), and were scanned within the first year following diagnosis. The study involved participants from 4 hospitals in France: Marseille, Rennes, Strasbourg, and Montpellier. **Table 1** provides details regarding the MRI scanners, participating centers, and characteristics of the study subjects.

---

[1] https://clinicaltrials.gov/ct2/show/NCT02117375

## 2.2. MRI Acquisition

Scans were acquired using Siemens 3T MRI scanners (Verio and Skyra). The following is a brief presentation of each MR modality that we processed.

For diffusion-weighted imaging (DWI), 30 images were acquired at $b$ = 900 s·mm$^{-2}$ with non-collinear gradient directions, along with six non-DWI ($b$ = 0) measurements and one non-DWI ($b$ = 0) with an opposite phase encoding direction (PED). This was repeated three times successively in order to increase the signal-to-noise ratio. Scans were performed in sagittal orientation with head-feet PED. The diffusion MRI pulse sequence used single-shot echo-planar imaging with parallel imaging (GRAPPA, acceleration factor 2). Sixteen slices were acquired without an inter-slice gap, with a resolution of 2×2×2 mm$^3$, an image matrix of 80×80, and a TR/TE of 3600/90 ms. The total acquisition time for the DWI sequence was approximately 7 minutes. Additionally, the protocol includes three anatomical references: a T1-weighted scan in sagittal orientation with a resolution of 1×1×1 mm$^3$, a TR/TE of 1800/2.79 ms, and field of view (FoV) of 250 mm; T2-weighted scan in sagittal orientation with a resolution of 0.7×0.7×2.75 mm$^3$, a TR/TE of 3000/68 ms, and an Fov of 260 mm; and an axial T2 scan with a resolution of 0.6×0.5×3 mm$^3$, a TR/TE of 4790/94 ms, and an FOV = 180 mm.

## 2.3. Analysis and processing pipeline

In this section, we present the processing pipeline, which takes raw images and extracts diffusion measures for each subject within the cervical spinal cord vertebral levels. To ensure that the results are immune to image artifacts and to identify problems during the processing pipeline, we conducted thorough quality control on the raw data and after each processing step. This stage is crucial for the quality of the analysis and the accuracy of the results.

### 2.3.1. Image preprocessing

Motion between DWIs was corrected using the method presented in (Xu et al., 2013) and implemented in the Spinal Cord Toolbox (SCT) (De Leener et al., 2017). Subsequently, dMRI data were corrected for susceptibility distortion using the Hyperelastic Susceptibility Artefact Correction (HySCO) method as implemented in the Statistical Parametric Mapping (SPM) toolbox and presented in (Ruthotto et al., 2012). HySCO demonstrated efficient performance for diffusion MRI of the spinal cord, as shown in (Snoussi et al., 2019). Using SCT, whole spinal cord segmentation was performed on T1-weighted images, as well as on the mean of DWIs ($b$ = 900 s·mm$^{-2}$) corrected for distortion. In some cases, this segmentation was ameliorated by adjusting parameters. We then manually identified two vertebral levels, C3 and T1, to fulfill the requirements for registering of T1-weighted data to the PAM50 template (De Leener et al., 2018).

### 2.3.2. Computation of diffusion-based metrics

The diffusion-weighted signal in white matter was modeled in the spinal cord using DTI (Basser et al., 1994) and the Ball-and-Stick model (Behrens et al., 2007). The DTI model assumes that the probability of water molecule displacement follows a zero-mean 3D Gaussian distribution. The diffusion tensor, directly related to the covariance matrix, is a 3×3 symmetric, positive-definite matrix. From its eigenvalue decomposition, we can extract rotation-invariant indices. We focused on radial diffusivity (RD), axial diffusivity (AD), mean diffusivity (MD), and fractional anisotropy (FA). Note that MD can be expressed as a combination of AD and RD. These DTI metrics were computed using the DIPY library (Garyfallidis et al., 2014).

In contrast to the DTI model, the Ball-and-Stick model is a two-compartment model, where each compartment provides a normalized MR signal, $S_1$ and $S_2$. These signal models correspond to intra- and extra-axonal diffusion, respectively. For the intra-axonal compartment, $S_1$ refers to signals from the water inside the axons where the diffusion is restricted. For the extra-axonal compartment, $S_2$ refers to signals originating from water outside the axons.

For the Ball-and-Stick model, the first compartment, $S_1$, is a *stick* (anisotropic component) with fiber direction **n** and diffusivity $d$ as parameters (Behrens et al., 2003). The *stick* compartment describes diffusion in an idealized cylinder with a zero radius. The signal for this component is:

$$S_1(d, n; b, G) = \exp\left(- bd(n \cdot G)^2\right) \quad (1)$$

where $b$ is the diffusion-weighting parameter and **G** is the gradient direction. The second compartment $S_2$, referred to as a *ball*, is an isotropic component with only the diffusivity $d_0$ as a parameter in its signal description:

$$S_2(b) = \exp\left(- bd_0\right) \quad (2)$$

In our implementation, we fixed λ2 and λ3 of the *stick* to $0.2 \times 10^{-3}\ mm^2/s$ and $d_0$ to $3.0 \times 10^{-3}\ mm^2/s$, which corresponds to the free diffusion coefficient of water. Fixing those values allows us to use this model on data with one non-zero b-value. The signal model is therefore:

$$S(d, n, f; b, G) = (1 - f)S_1(d, n; b, G) + fS_2(b)$$

The parameters of interest that we extracted from the Ball-and-Stick model are $f$, the free water weight (FWW), and $d$, the stick axial diffusivity (Stick-AD) (Snoussi et al., 2023).

### 2.3.3. Quantification of metrics per vertebral level

To calculate the mean of the presented metrics for each vertebral level, we followed a processing pipeline to align the labels defined in the PAM50 template (De Leener et al., 2018) with the native DWI space of each subject.

First, the T1-weighted anatomical image was registered to the PAM50-T1 spinal cord template (De Leener et al., 2018); generating a forward and inverse warping field between them. Next, the PAM50-T1 template (De Leener et al., 2018) was registered to the mean DWI using the inverse warping field from the previous registration as an initial warping field. This registration was performed using T1-weighted images instead of T2-weighted images because their isotropic resolution made the registration more effective.

Thus, alignment with the template provides a robust definition of the inter-vertebral levels for the spine. This allows for computation of average metrics in the spinal cord using the atlas-based approach introduced in (Lévy et al., 2015). As a result, we can quantify diffusion-based metrics averaged for each vertebral level in the cervical part. Specifically, for scalar metrics, we quantify them only in white matter according to the PAM50 template (Snoussi et al., 2022). The processing pipeline is summarized in **Figure 1**.

### 2.3.4. Ground truth: Segmentation of MS lesion

For the 53 MS patients, MS lesions were manually segmented by nine raters, including radiologists and experienced readers, as described in (Eden et al., 2019). Briefly, MS lesions were segmented using both axial T2 and sagittal T2-weighted images with ITK-SNAP Toolbox 3.6.0 (Yushkevich et al., 2006). From these lesion masks, we computed for each vertebral level: (i) the number of MS lesions within the vertebral level, (ii) the total lesion volume normalized by the volume of the corresponding vertebral level. In **Figure 2**, red bars represent the quantity and distribution of lesion volumes in the [C1-C7] region, referencing C1, C2, .., C7 levels.

## 2.4. Statistical analysis

From the processing pipeline described above, we obtained six diffusion-based metrics: FWW, Stick-AD, AD, FA, MD, and RD. All metrics were quantified and averaged for each vertebral level of every subject in our cohort. In this section, we present our proposed statistical analysis study comparing MS patients and healthy volunteers, examining each metric separately.

### 2.4.1. Pairwise comparison

Our cohort consists of 29 healthy volunteers and 53 MS patients, with 139 segmented lesions of various volumes distributed across cervical vertebral levels. This sample size is relatively small compared to the number of extracted metrics and the volume of the vertebral level in which we computed the average of the proposed metrics. Additionally, it is common for such statistical analysis to be performed for each vertebral level separately due to the anatomical variety within the cervical spinal cord.

Therefore, we investigated the potential for pooling data from multiple vertebral levels for the purpose of increasing our analysis's statistical power. To do this, we performed a two-way analysis of variance between vertebral levels for each metric to illustrate the interaction term between them. We compared all pairs of vertebral levels of one subject with each level of all the other subjects. This test reveals the degree to which one subject is differentially effective at each vertebral level of a second subject. This test was performed using estimated marginal means, sometimes called least-squares means, which are predictions from a linear model over a reference grid or marginal averages thereof.

**Figure 3** graphically illustrates this comparison for the six-diffusion metrics, four from DTI and two from the Ball-and-Stick. These metrics were computed using data solely from the 29 healthy volunteers. **Figure 3** summarizes intervals of vertebral levels where no significant difference exists. We observed that the [C2-C4] region shows no significant difference for all metrics. This finding suggests the possibility of combining and pooling metrics quantified in C2, C3, and C4 vertebral levels, thereby increasing the available data for our statistical analysis. **Figure 2** displays the quantity and distribution of lesion volume within the [C2-C4] region in comparison to the [C1-C7] region.

### 2.4.2. Unpaired t-test between healthy volunteers and MS patients

For the data of the [C2-C4] region, we performed Welch's *t*-test between healthy volunteers and MS patients. This *t*-test is an adaptation of Student's *t*-test and provides more reliable results when the two samples have unequal variances and/or unequal sample sizes. It is commonly referred to as *unpaired* or *independent samples t*-tests. In this statistic test, for each metric $m_i$, we have the following:

$$V_i = [h_1[m_i(c_2), m_i(c_3), m_i(c_4)], ..., h_{29}[m_i(c_2), m_i(c_3), m_i(c_4)]]$$
$$NAWM_i = [p_1[m_i(c_j)], ..., p_{53}[m_i(c_j)]]; \; c_{j \in (2,3,4)}$$
$$MS_i(thr) = [p_1[m_i(c_j)], ..., p_{53}[m_i(c_j)]]; \; c_{j \in (2,3,4)}$$

where, $h$ refers to healthy volunteer, $p$ refers to MS patient, $m_i$ is the chosen metric with its index $i \in \{1, .., 6\}$, $c_2$, $c_3$ and $c_4$ are the vertebral level, $thr$ is the threshold of the percentage of the lesion's volume, $V_i$: 87 (29×3) vertebral levels from 29 healthy controls, $NAWM_i$: 86 vertebral levels without detected lesion, i.e. as normal-appearing white matter ($NAWM$), $MS_i(threshold\%)$: vertebral levels possess lesion with volume superior to $thr\%$ of the corresponding vertebral level volume. To maintain reasonable statistical power in this *t*-test, we chose only two thresholds: $thr = 5\%$ and $thr = 10\%$. Consequently, 36 and 24 vertebral levels in our cohort possess lesion volume greater than $thr = 5\%$ and $thr = 10\%$, respectively.

### 2.4.3. Machine learning for the detection of MS lesions

In this section, we propose to utilize and evaluate diffusion MRI data to detect the presence of a lesion automatically. Throughout this part, we will assess classification results using the area under the curve (AUC) of the receiver operating characteristic (ROC) curve.

### 2.4.4. Multivariate classification using diffusion metrics

*Building linear discriminant analysis classifier:* Based on a selection of metrics extracted from diffusion data, we construct a classifier that combines this set of features using the Linear Discriminant Analysis (LDA) method *(Ripley, 2002)*. LDA is a technique employed to identify a linear combination of features that separates or characterizes two or more classes of objects. The resulting combination can be used as a linear classifier.

The whole experience setup is summarized in **Algorithm 1**. So, the data vector $X_{comb}$ is constructed as the following:

$$X_{comb}(thr) = \left[\left[X_i(thr)\right], .., \left[X_j(thr)\right]\right] \quad (4)$$

Where $i$ and $j$ are the index of the chosen metrics and $thr$ is the threshold of the lesion volume.

Note that depending on the threshold of the lesion's volume, the number of vertebral levels in the patient group may vary. In fact, $X_{comb}$ contains vertebral levels of healthy volunteers (29×3) and vertebral levels of MS patients with lesions. **Figure 2** shows the count of vertebral levels having lesions for different cumulative percentage threshold for lesion volume in [C2-C4] region. We report the mean and standard deviation of ROC AUC for 1,000 splits of the dataset into training and testing parts, representing 67% and 33% of the original dataset, respectively. This figure provides an idea about the sample size of training and testing datasets used in the subsequent analysis.

*Selecting a subset of measures:* As mentioned earlier, due to the relatively sample size, we need to reduce our linear classifier's degrees of freedom by choosing a subset of metrics. Our goal is to select a subset of diffusion-based metrics that provide complementary information. To accomplish this, we first calculated the normalized covariance matrix for all metrics in the [C2-C4] region on healthy volunteers $V$ and $MS(10\%)$ as shown in ***Figure 4***. Dark blue squares indicate a strong correlation between two metrics, while yellow squares signify no relationship between them. Based on these correlations, we propose combinations of metrics to be studied in **Table 2**. A critical remark here is that the normalized covariance matrix presented in **Figure 4** reveals a difference in correlation between some diffusion-based metrics derived from healthy and affected vertebral levels. Particularly, the correlation between FWW and RD decreases when metrics are quantified in MS vertebral levels. We will focus on metrics with high potential because they exhibit good classification performance and present significant differences between MS patients and controls (see subsection *2.4.2*).

**Algorithm 1.** ROC AUC for a combination of metrics

---

I. Fix *threshold* of $MS$ lesion from $\{0.02, 0.04, .., 0.20\}$.

II. Construct the data vector $X_{comb}$ and its label vector $Y_{comb}$ (0 to healthy volunteers and 1 to $MS$ patients).

III. Standardize $X_{comb}$ to get $X_{scaled}$ by centering to the mean and component wise scale to unit variance.

IV. Split $X_{scaled}$ 1000 consecutive times in different $X_{train}$ (67%) and $X_{test}$ (33%) with their corresponding $Y_{train}$ and $Y_{test}$

  IV.1. Fit LDA using $X_{train}$ and $Y_{train}$

  IV.2. Using the fitted LDA model, Predict confidence score on $X_{test}$ to obtain $Y_{LDA}$

  IV.3. Compute ROC AUC score between $Y_{test}$ and $Y_{LDA}$

V. Calculate the mean and variance of ROC AUC scores which is computed in 1000 consecutive times.

---

## 3. Results

### 3.1. Unpaired *t*-test between healthy volunteers and MS patients

**Table 3** presents the mean, standard deviation, and p-value of each diffusion metric *i* for healthy volunteers' data $V_i NAWM_i \text{[OBJ]} MS_i(5\%) \text{[OBJ]} MS_i(10\%) \text{[OBJ]}$ as introduced in subsection 2.4.2.

FWW significantly increases in $MS$ patients, regardless of the presence and volume of lesions. For the second component of the Ball-and-Stick model, Stick-AD, there is a significant decrease in MS patients. FA demonstrates a significant reduction in MS patients with lesion volume exceeding 5% and 10% of the corresponding vertebral level volume. MD and RD increase significantly in MS patients with lesion volume greater than 5% and 10%. However, no detectable difference exists between values for healthy volunteers and MS patients for AD. It is important to note that for this unpaired *t*-test between healthy and affected vertebral levels in the [C2-C4] region, we fixed the threshold of the lesion's volume as a compromise with the size of available data. FWW, FA, MD, and RD of MS patients still show significant differences for various thresholds until the lesion's volume comprises 22% of the corresponding vertebral level, but Stick-AD has a p-value < 0.05 only until 12%.

### 3.2. Multivariate classification using diffusion metrics

**Figure 5** presents the mean and variance of the ROC AUC for each combination predicted by LDA, as introduced in **Table 2** superimposed by the ROC AUC mean of each metric used in the combination. This superposition is useful because it shows whether utilizing multipe metrics improves upon using each metric separately. The combinations presented in this part are selected and derived mainly after considering the covariance matrix, the unpaired t-test results, and their ROC AUC scores.

*Subset of 2 metrics*: When combining FWW and Stick-AD metrics, the ROC AUC mean score for separating the vertebral level of controls and MS patients with lesions is better than using each metric independently. For [FWW, FA] and [FA, MD], the combination is slightly better, as the ROC AUC score of each metric is still within the variance margin of the ROC AUC score of the combination. However, for [FWW, RD], the ROC AUC mean is similar to or close to the ROC AUC of the RD metric. When the lesion's volume greater than 10%, MS(thr > 10%), the best classification scores are approximately in [0.83, 0.87] using [FWW, FA] and [Stick-AD, FWW].

*Subset of 3 metrics*: **Figure 5** also displays the mean and variance of the ROC AUC score for combinations of 3 metrics: [RD, MD, FA] and [FWW, MD, Stick-AD]. For these combinations, we observe that the ROC AUC mean of the combination is better than the ROC AUC score of each metric independently. For MS(thr >10%), [FWW, MD, Stick-AD] has ROC AUC mean in [0.82, 0.86] and [RD, MD, FA] in [0.86, 0.90], which is an interesting result.

*Subset of 4 metrics*: Additionally, we present two combinations of 4 metrics: [RD, FWW, FA, MD] and [FWW, Stick-AD, MD, RD]. For MS(thr > 10%), [RD, FWW, FA, MD] has an ROC AUC mean in [0.84,0.86] and [FWW, Stick-AD, MD, RD] has an ROC AUC mean in [0.87, 0.91].

In summary, we can deduce that among all combinations, [RD, MD, FA] and [FWW, Stick-AD, MD, RD], which are overlaid in **Figure 6**, yield the best prediction scores for distinguishing between healthy volunteers and MS patients with a lesion. The minimum variance margin for these subsets is close to or slightly better than the best ROC AUC score of the independent RD or FA metrics when the lesion's volume is greater than 10%.

### 4. Discussion

In this study, we aimed to investigate the sensitivity of diffusion MRI for identifying MS lesions in the cervical spinal cord. We established a pipeline that incorporates several image processing techniques and an atlas-based approach to calculate the average of diffusion MRI metrics for each vertebral level in the cervical spinal cord.

We derived diffusion measurements from DTI and Ball-and-Stick models, followed by a statistical analysis to evaluate their sensitivity associated with the presence of MS lesions within the same vertebral level. This analysis included an unpaired t-test and multivariate classification. Our spinal cord cohort was acquired and collected from multiple clinical sites.

In our work, we conducted a two-way analysis of variance between vertebral levels for each metric to illustrate and demonstrate the interaction term between them. Our results indicated no significant inter-difference between C2, C3, and C4 vertebral levels for all six diffusion metrics, as illustrated in **Figure 3**. In fact, pooling [C2-C5] instead of [C2-C4] was possible for FWW, Stick-AD, FA, MD, and RD. However, for the MD metric, we observed a p-value of 0.065 between C3 and C5, which was close to being significant. As a result, we preferred to focus our statistical analysis on the [C2-C4] interval.

We discovered that FWW, Stick-AD, FA, MD, and RD exhibited significant differences between healthy volunteers and MS patients within the [C2-C4] region of the cervical spinal cord. Although previous studies have demonstrated the involvement of FA, MD, and RD in the spinal cord (Agosta, Absinta, et al., 2007; Agosta, Pagani, et al., 2007; Valsasina et al., 2005; von Meyenburg et al., 2013), our work offers several important additions and contributions. Firstly, our dataset is larger, which enhances the robustness of our results. Secondly, our MS lesion segmentation was performed by nine raters, including radiologists and experienced readers, which ensured high accuracy. Thirdly, we quantified the diffusion measures within each vertebral level using an atlas-based approach. Fourthly, we performed a strict quality check of each step of the pipeline to ensure the accuracy of our results.

In addition to our study using DTI metrics, we also investigated the sensitivity of FWW and Stick-AD, two measurements derived from the Ball-and-Stick model, for detecting lesions in MS patients. Our findings demonstrated that the Ball-and-Stick multicompartment model offers valuable insights into the tissue microstructure in lesioned regions of MS patients. This model, unlike traditional DTI, is capable of capturing more complex tissue architecture, thus providing a deeper understanding of the underlying pathological processes in MS. In our study, we observed that axial diffusivity (AD) did not exhibit any significant sensitivity between healthy volunteers and MS patients. However, Stick-AD, a metric derived from the Ball-and-Stick model, showed significant differences between the two groups. This suggests that Stick-AD might be better suited for detecting microstructural changes in MS lesions compared to conventional AD. Moreover, we found that FWW, the free water-weighted compartment of the Ball-and-Stick model, which is not obtainable from traditional DTI, displayed high accuracy in discriminating lesioned vertebral levels from healthy ones. This indicates that FWW may offer unique and clinically relevant information regarding tissue microstructure in MS patients. It is important to note that our results were obtained using a regularized version of the Ball-and-Stick model, in which the ball's diameter and the second and the third eigenvalues of the stick were fixed manually. We based this manual adjustment on the state-of-the-art methods and Anima-Public Software (https://anima.irisa.fr). Despite using a regularized model, our findings still highlight the potential advantages of the Ball-and-Stick model in understanding MS pathology.

Therefore, our findings underscore the importance of employing acquisition protocols with multiple b-values that are designed to enable richer multi-compartment models (Scherrer & Warfield, 2012). To our knowledge, neurite orientation dispersion and density imaging (NODDI) is the only multi-compartmental diffusion model used for assessing microstructure and characterizing abnormalities in the spinal cord of MS patients (By et al., 2017). But this study only analyzed a single slice of the cervical spine to evaluate the sensitivity and feasibility of NODDI in MS patients.

Furthermore, our study made a significant contribution by exploring a multivariate learning approach to automatically detect the presence of an MS lesion using diffusion MRI data. We trained a linear classifier using LDA, based on a selection of metrics extracted from diffusion MRI with limited cross-correlation. We discovered that combining certain metrics improved the prediction accuracy for the presence of MS lesions, outperforming the use of individual metrics, as illustrated in **Figure 5** and **Figure 6**. Consequently, we determined that combining three metrics [FA, RD, MD] and four metrics [FWW, MD, Stick-AD, RD] resulted in better ROC AUC scores when differentiating between healthy volunteers and MS patients with lesions. For MS patients with a lesion volume greater than 10%, the [FA, RD, MD] combination yielded a mean ROC AUC score in the range of [0.86, 0.90], while the [FWW, Stick-AD, MD, RD] combination had a mean ROC AUC score in the range of [0.87, 0.91]. These prediction score intervals indicate that the classification accuracy is good and superior to using individual metrics independently.

## 5. Conclusion

We demonstrated the sensitivity of DTI and Ball-and-Stick reconstruction models to underlying microstructure changes in MS within the context of a multicenter study. A multi-compartment model, Ball-and-Stick, provides novel information about the tissue microstructure in lesioned regions of MS patients, offering potential improvements over traditional DTI methods. Our study reveals the significance of Stick-AD and the value of FWW in discriminating lesioned vertebral levels, even when using a regularized version of the model. Furthermore, we identified that combining several diffusion metrics together enabled us to distinguish between lesioned and non-lesioned vertebral levels with higher accuracy. We showed that selecting a subset of metrics, [FA, RD, MD] and [FWW, MD, Stick-AD, RD], which offer complementary information, significantly increased the prediction accuracy for the presence of MS lesions in the cervical spinal cord. Our study provides novel insights and highlights the potential of multivariate statistical analysis for assessing tissue microstructure and pathology.

**Table 1.** Demographic and clinical information about the participated clinical sites and for all participants, healthy volunteers and MS patients in our cohort.

| Center | Center 1 | Center 2 | Center 3 | Center 4 | TOTAL |
|---|---|---|---|---|---|
| 3T MRI | Verio | Verio | Verio | Skyra | - |
| Volunteers | 4 | 18 | 3 | 4 | **29** |
| Gender | F/3M | 10F/8M | 3F | 4F | **18F/11M** |
| Mean age(year) | 34.0±4.74 | 32.61±7.97 | 34.67±5.25 | 31.25±5.67 | **32.83±7.13** |
| Mean weight(kg) | 72.5 ± 6.7 | 65.4 ± 11.4 | 65.0 ± 7.5 | 56.0 ± 4.1 | **65.0 ± 10.7** |
| Mean height(m) | 1.75 ± 0.03 | 1.72 ± 0.09 | 1.66 ± 0.06 | 1.64 ± 0.04 | **1.71 ± 0.08** |
| MS Patients | 6 | 35 | 5 | 7 | **53** |
| Gender | 4F/2M | 22F/13M | 3F/2M | 5F/2M | **34F/19M** |
| Mean age(year) | 34.17±7.90 | 31.74±6.08 | 33.80±5.91 | 34.57±4.78 | **32.58±6.25** |
| Mean weight(kg) | 69.7 ± 10.5 | 67.9 ± 13.9 | 65.6 ± 6.7 | 69.1 ± 13.6 | **68.1 ± 13.0** |
| Mean size(m) | 1.68 ± 0.08 | 1.70 ± 0.09 | 1.71 ± 0.07 | 1.68 ± 0.06 | **1.70 ± 0.09** |
| | | | | | **82 (52F/30M)** |

**Table 2.** Proposed combinations of 2, 3 and 4 metrics to be studied. FWW: free water weight, Stick-AD: stick axial diffusivity, AD: axial diffusivity, FA: fractional anisotropy, MD: mean diffusivity, and RD: radial diffusivity.

| Subsets of 2 Metrics | Subsets of 3 and 4 Metrics |
|---|---|
| FWW & RD | FA & MD & RD |
| FWW & FA | FWW & MD & Stick-AD |
| FWW & Stick-AD | FWW & MD & Stick-AD & RD |
| FA & MD | FWW & MD & FA & RD |

**Table 3.** For C2, C3, and C4 levels, the mean and STD of each metric for healthy volunteers, for MS patients with or without lesions, and MS patients with lesions >5% and >10%. Dark green means that p-value is inferior to 0.01: there is a significant difference between healthy volunteers and MS patients. Weak green means that p-value shows significant difference but 0.01 < p-value < 0.05. n represents the number of vertebral level data available in the [C2-C4] region. Vi: vertebral levels from 29 healthy controls, n: number of vertebral levels from 29 healthy controls, NAWM: normal-appearing white matter, STD: standard deviation, FWW: free water weight, Stick-AD: stick axial diffusivity, AD: axial diffusivity, FA: fractional anisotropy, MD: mean diffusivity, and RD: radial diffusivity.

| Data | Healthy Volunteers | | MS patients | | | | | | | | |
|---|---|---|---|---|---|---|---|---|---|---|---|
| | $V_i$ (n=87) | | $NAWM_i$ (n=86) | | | $MS_i$ (5%) (n=36) | | | $MS_i$ (10%) (n=24) | | |
| Metric | Mean | STD | Mean | STD | p-value | Mean | STD | p-value | Mean | STD | p-value |
| FWW (mm$^2$/s) | 0.1594 | 0.0431 | 0.1774 | 0.0672 | 0.0398 | 0.2076 | 0.0735 | 0.0007 | 0.2087 | 0.0641 | 0.0016 |
| Stick-AD (10$^3$ mm$^2$/s) | 1.1419 | 0.2759 | 1.0994 | 0.2640 | 0.3048 | 1.0356 | 0.2398 | 0.0378 | 1.0221 | 0.2524 | 0.0401 |
| AD (10$^3$ mm$^2$/s) | 1.6516 | 0.2105 | 1.6090 | 0.2571 | 0.2381 | 1.6748 | 0.2101 | 0.5844 | 1.7020 | .1500 | 0.1978 |
| FA | 0.6899 | 0.0800 | 0.6774 | 0.0941 | 0.3490 | 0.6150 | 0.0908 | 0.0001 | 0.6098 | 0.0667 | 1e-5 |
| MD (10$^3$ mm$^2$/s) | 0.8370 | 0.1190 | 0.8330 | 0.1882 | 0.8685 | 0.9234 | 0.1656 | 0.0071 | 0.9378 | 0.0931 | 8e-5 |
| RD (10$^3$ mm$^2$/s) | 0.4297 | 0.1302 | 0.4450 | 0.1838 | 0.5315 | 0.5477 | 0.1774 | 0.0008 | 0.5556 | 0.1095 | 3e-5 |

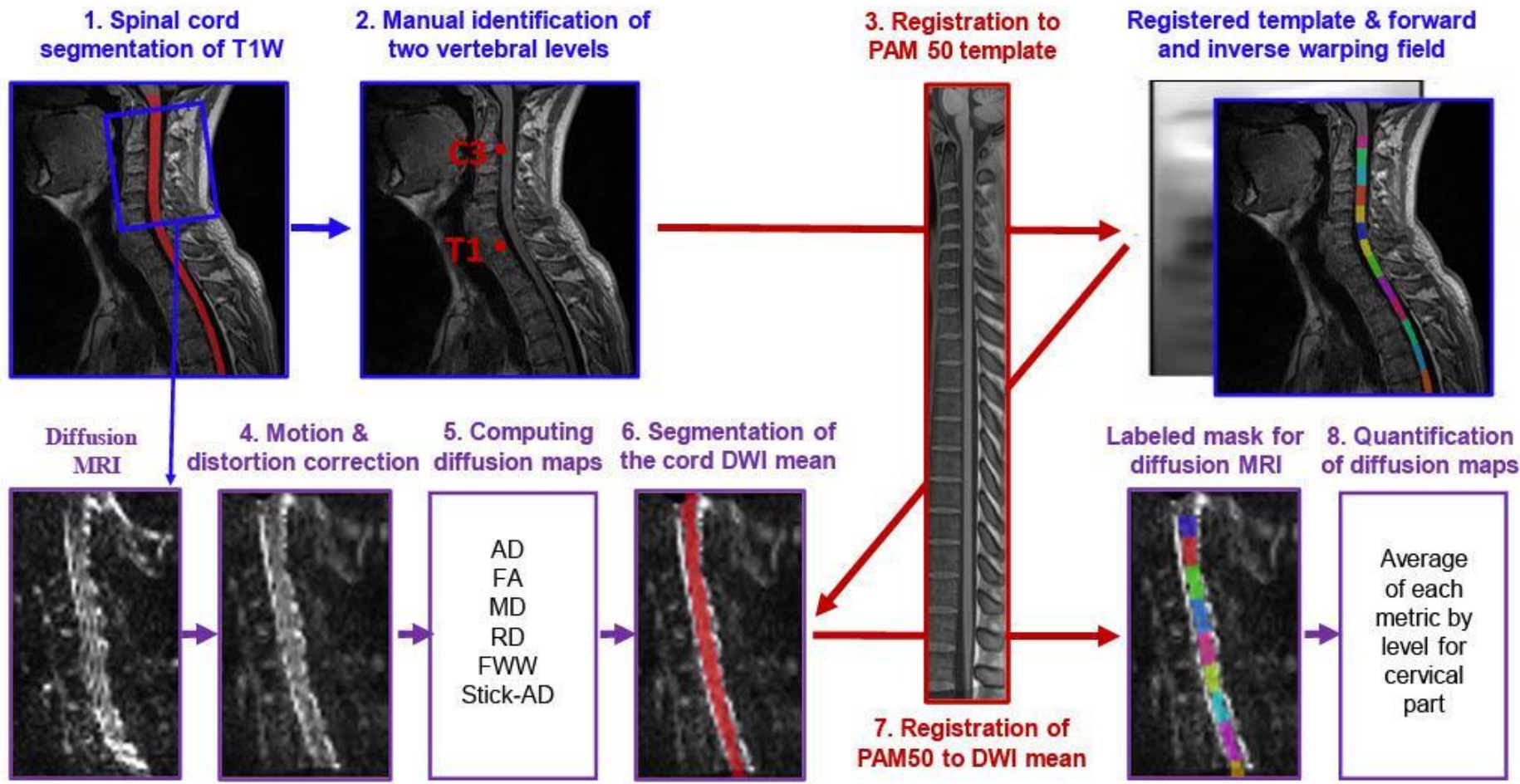

**Figure 1.** Illustration of the analysis pipeline, which includes the following steps: (1) Segmentation of the spinal cord on T1W. (2) Manual identification of two vertebral levels. (3) Registration of T1W image to the PAM50 template. (4) Motion and distortion correction of dMRI data. (5) Computation of DTI and Ball&Stick metrics. (6) Segmentation of the spinal cord using the mean of the DWI data. (7) Registration of the PAM50-T1W registered to DWI mean data using the inverse warping field from the previous registration as an initial warping field. (8) Quantification of metrics within each vertebral level of the cervical part. T1W: T1

Weighted, PAM50: template and atlas of the white and gray matter spinal cord, DWI: Diffusion Weighted Imaging, AD: axial diffusivity, FA: fractional anisotropy, MD: mean diffusivity, RD: radial diffusivity, FWW: free water weight, and Stick-AD: stick axial diffusivity. The unit of FWW, Stick-AD, AD, MD and RD is mm$^2$/s.

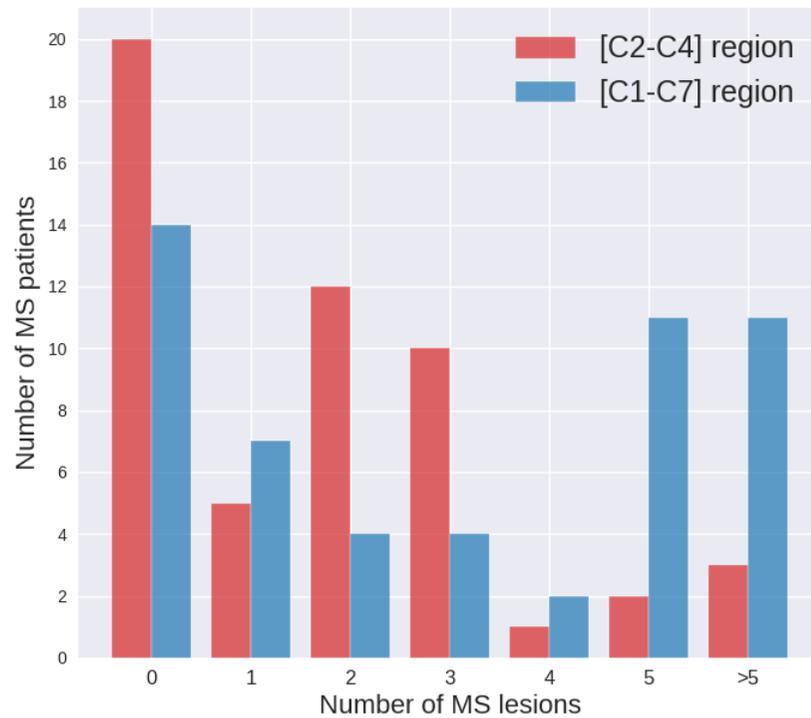 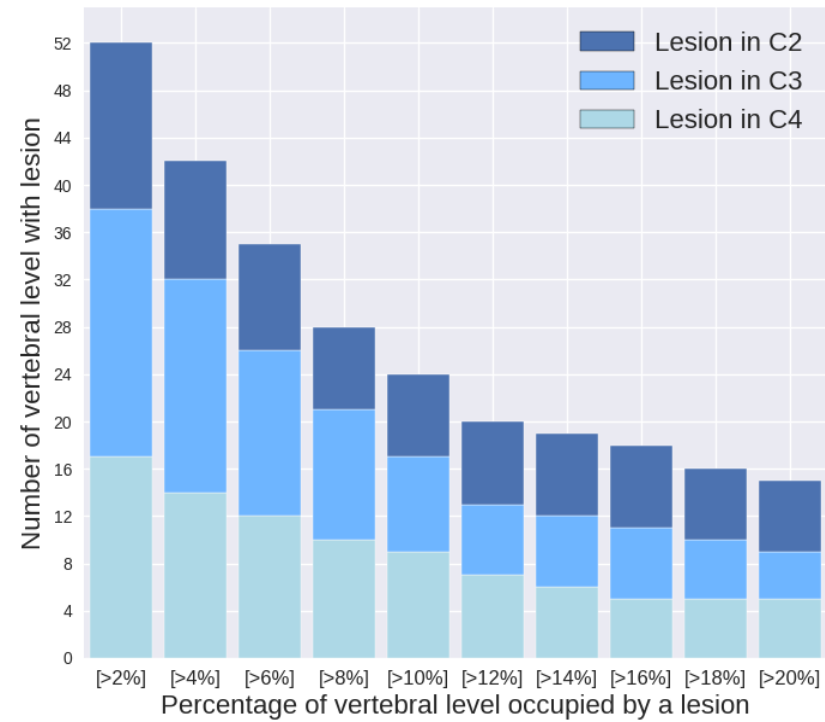

**Figure 2.** Left subfigure: Distribution of MS within our MS cohorts. Right subfigure: Distribution of lesion's volume in [C2-C4] region. The y-axis shows the number of segmented lesions in [C2-C4] regions, while the x-axis represents the threshold percentage for lesion volume. Lesion's volume is the part of the vertebral volume occupied by a lesion. C1, ..C7 refer to the cervical vertebral levels.

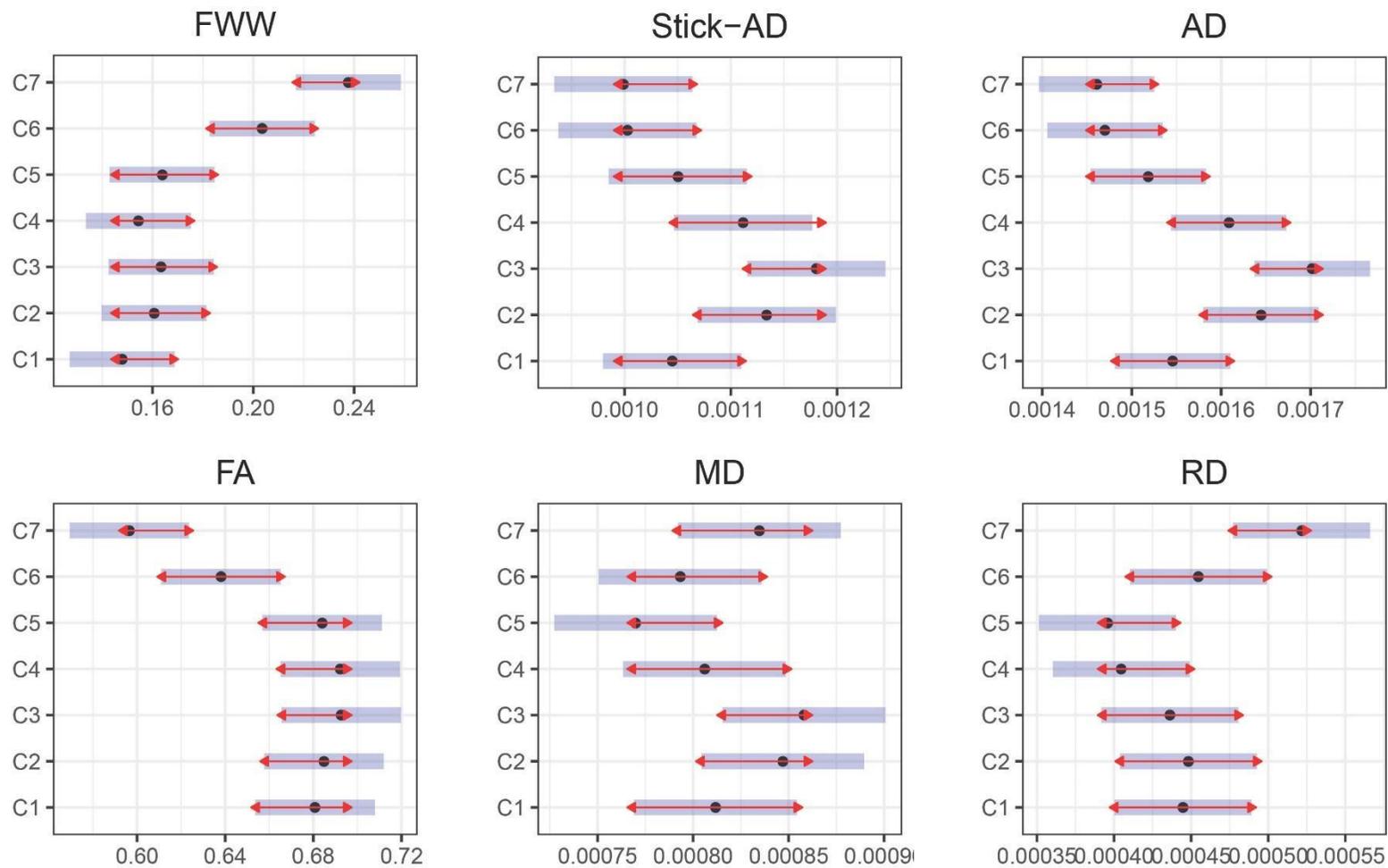

**Figure 3.** Estimated marginal means (x-axis) for each metric in cervical vertebral levels (y-axis) for healthy volunteers data. The blue bars are confidence intervals for the estimated marginal means, and the red arrows are for the comparisons among them. If an arrow from one level overlaps an arrow from another level, the difference is not significant (*p*-value >0.05). Else, the difference is significant

(p-value<0.05). FWW: free water weight, Stick-AD: stick axial diffusivity, AD: axial diffusivity, FA: fractional anisotropy, MD: mean diffusivity, and RD: radial diffusivity. The unit of FWW, Stick-AD, AD, MD and RD is mm²/s.

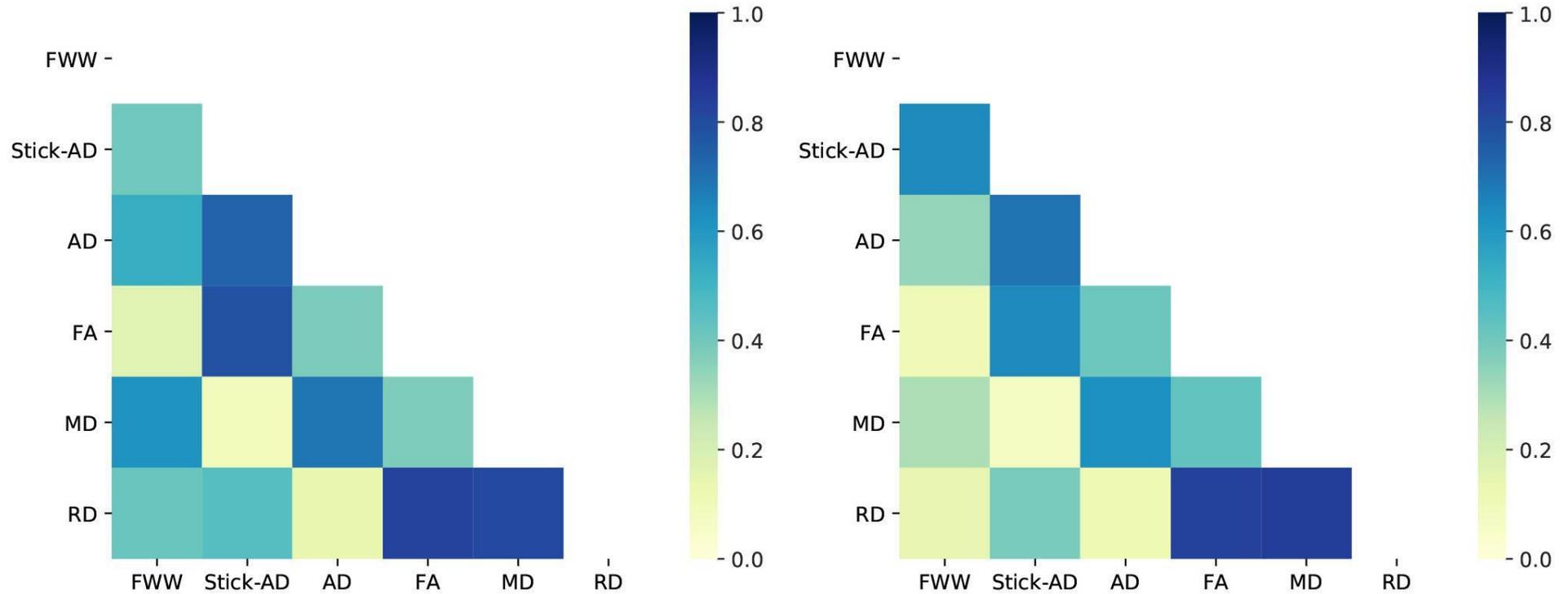

**Figure 4.** Normalized covariance matrix of metrics in [C2-C4] levels for healthy volunteers *V* (left), MS patients *MS*(10%) (right). The dark blue square shows a strong correlation between the two metrics, and the white square indicates no relationship between them. FWW: free water weight, Stick-AD: stick axial diffusivity, AD: axial diffusivity, FA: fractional anisotropy, MD: mean diffusivity, and RD: radial diffusivity. The unit of FWW, Stick-AD, AD, MD and RD is mm²/s.

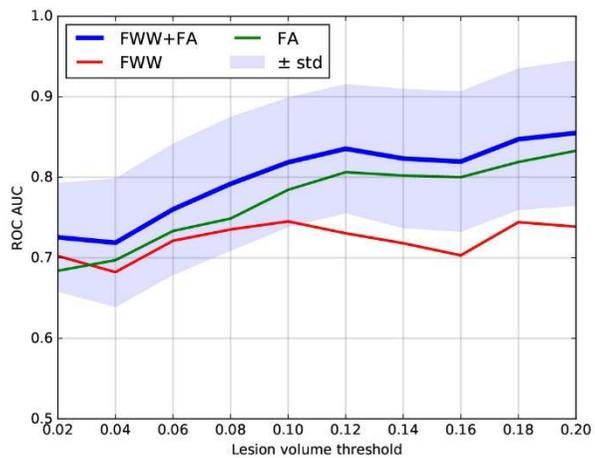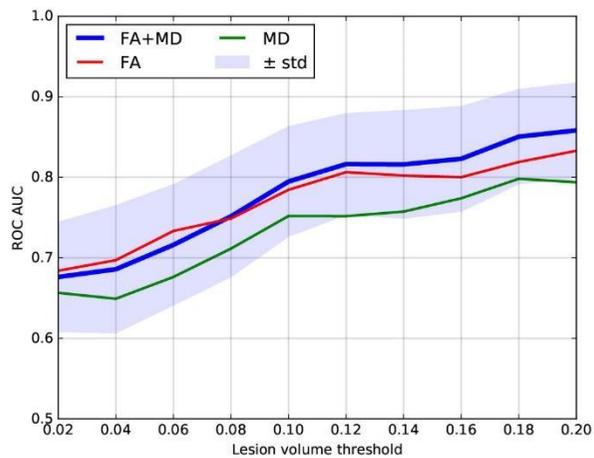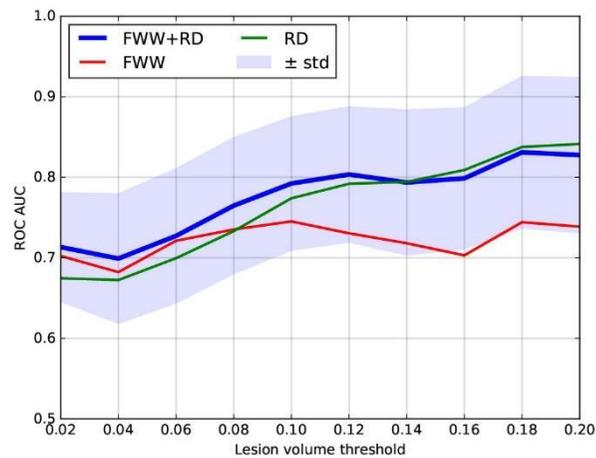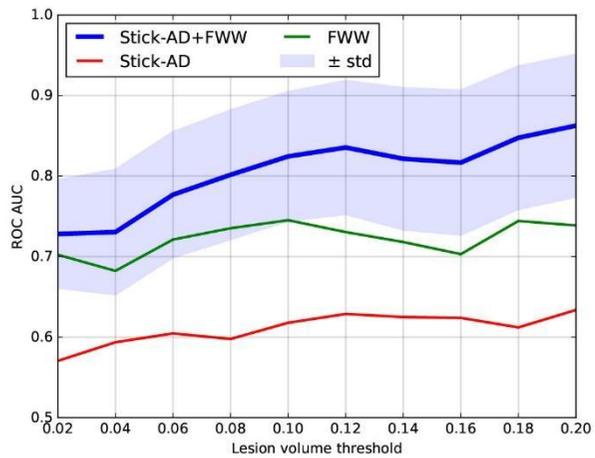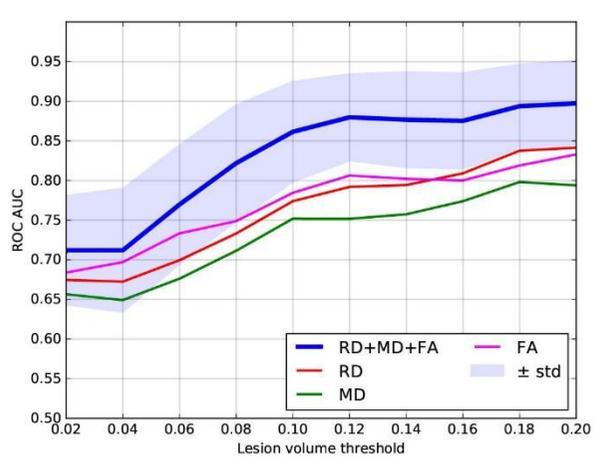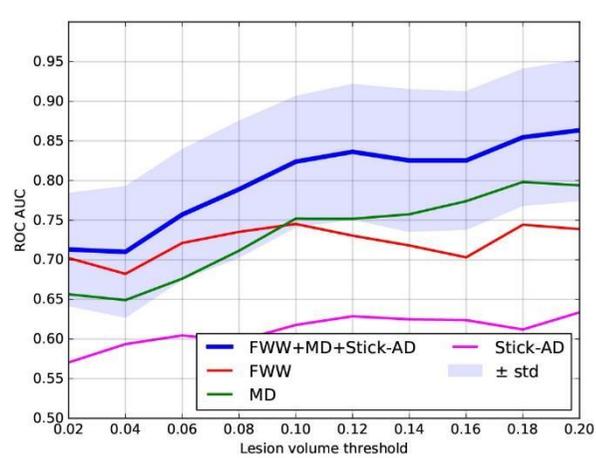

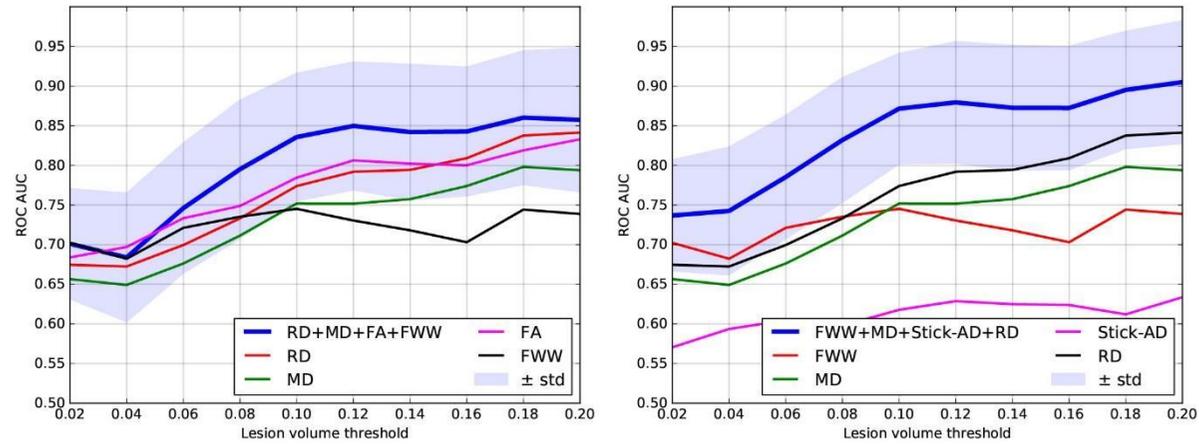

**Figure 5.** ROC AUC predicted by LDA as a function of lesion volume percentage in the corresponding vertebral level for each combinations of metrics, and ROC AUC of native metrics between MS patients and controls. ROC AUC: the area under the curve (AUC) of the receiver operating characteristic curve (ROC), STD: standard deviation, FWW: free water weight, Stick-AD: stick axial diffusivity, AD: axial diffusivity, FA: fractional anisotropy, MD: mean diffusivity, and RD: radial diffusivity.

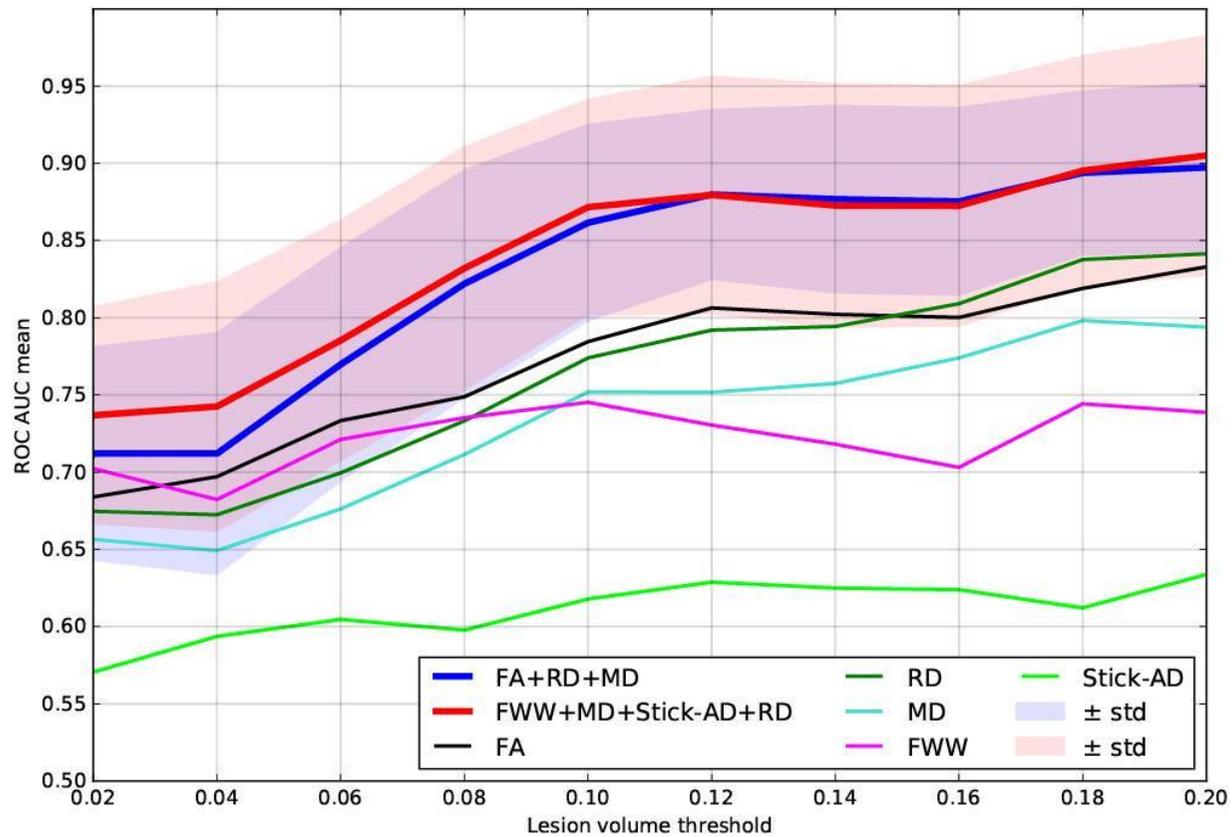

**Figure 6.** Overlays of ROC AUC score predicted by LDA as a function of lesion volume percentage in the corresponding vertebral level for the best combinations: [FA, MD, RD] and [FWW, MD, Stick-AD, RD]. ROC AUC: the area under the curve (AUC) of the receiver operating characteristic curve (ROC), STD: standard deviation, FWW: free water weight, Stick-AD: stick axial diffusivity, AD: axial diffusivity, FA: fractional anisotropy, MD: mean diffusivity, and RD: radial diffusivity.